\documentstyle[epsfig]{article}

\begin{document}

\title{Analyzing power in nucleon-deuteron scattering and three-nucleon forces}
\author{S. Ishikawa\\
Department of Physics, Hosei University,\\
 2-17-1 Fujimi, 102-8160 Tokyo, Japan\\ 
and \\
Frontier Research Center for Computational Sciences, \\
Science University of Tokyo, \\
2641 Yamazaki, Noda, 278-8510 Chiba, Japan
}

\maketitle

\begin{abstract}
Three-nucleon forces have been considered to be one possibility to 
resolve the well known discrepancy between experimental values and 
theoretical calculations of the nucleon analyzing power in low energy 
nucleon-deuteron scattering.
In this paper, we investigate possible effects of two-pion exchange 
three-nucleon forces on the analyzing power and the differential cross section.
We found that the reason for different effects on the analyzing power by 
different three-nucleon forces found in previous calculations is related to 
the existence of the contact term.
Effects of some variations of two-pion exchange three-nucleon forces are 
investigated.
Also, an expression for the measure of the nucleon analyzing power with 
quartet P-wave phase shifts is presented.
\end{abstract}
Differential cross sections for nucleon-deuteron elastic scattering have 
peaks at forward and backward scattering angles and a minimum at a c.m. 
scattering angle of, e.g., $\theta \sim 105^\circ$ at $E_{Lab}^N$ = 3 MeV.
Around the cross section minimum angle, some observables calculated with 
realistic nucleon-nucleon (NN) potentials are known to deviate 
systematically from experimental values \cite{Gl96}.
The nucleon analyzing power $A_y(\theta)$ for energies below  
$\approx$ 30 MeV has exhibited a notable discrepancy \cite{Ko87,Wi88}, 
which is referred to as the $A_y(\theta)$ puzzle.
E.g., in the neutron-deuteron (n-d) elastic scattering at $E_{Lab}^n$ = 3 
MeV, experimental $A_y(\theta)$ has a maximum value at 
$\theta \sim 105^\circ$ \cite{Mc94}, while theoretical calculations with 
modern realistic NN potentials \cite{St94,Wi95,Ma96} undershoot the value by 
about 30 \%.
The three-nucleon (3N) system has been considered as a good testing ground 
for the NN interaction.
The discrepancy between the experimental and calculated $A_y(\theta)$ may 
show that there is room for improvement of modern NN potentials.
Actually, it was pointed out that changes in ${}^3P_J$ NN forces or 
the spin-orbit component of a potential cause a dramatic increase in 
$A_y(\theta)$ \cite{Wi91,Ta91,To98,Do98}.
However constraint from NN observables made it difficult to obtain 
reasonable changes in the NN potential to resolve the $A_y(\theta)$ 
puzzle \cite{Wi92,Hu98,To98b}.

Another possibility for resolving the $A_y(\theta)$ puzzle is the 
introduction of a three-nucleon force (3NF) into the nuclear Hamiltonian.
It is well known that most realistic NN forces underbind the triton, and a 
3NF based on exchange of two pions among the three nucleons (2$\pi$E-3NF) can 
explain the needed attraction.
So far, several 2$\pi$E-3NF models have been proposed, among which the 
Tucson-Melbourne (TM) 3NF \cite{Co79} and the Brazil 
(the earlier version, BR${}^\prime$ \cite{Co83}, 
and the latter version, BR \cite{Ro86}) 
3NF have been used for 3N calculations.
Although these 3NF models were made based on different ideas in constructing 
off-shell $\pi N$ scattering amplitudes which are important ingredients in 
2$\pi$E-3NF, the resulting potentials have essentially the same form with 
slightly different parameters.
It is reported that with introducing the TM-3NF or BR-3NF, the calculated 
$A_y(\theta)$ decreases, which means that the discrepancy with the 
experimental value is enhanced \cite{Wi94,Is94}.
On the other hand, the calculations with the BR${}^\prime$-3NF or 
another 3NF model, the Urbana (UR) 3NF, are reported to improve $A_y(\theta)$ 
slightly  \cite{Ki95,Ki96}.
The UR-3NF is based on  the $\Delta$-mediated two-pion exchange 
diagram \cite{Fu57}, which is a part of diagrams included in TM-3NF and BR-3NF.
The discrepancy of the effects on $A_y(\theta)$ should arise from a 
structure difference between TM/BR-3NF and BR${}^\prime$/UR-3NF. 
In this paper, we study effects of the 2$\pi$E-3NF on $A_y(\theta)$ carefully 
and investigate the possibility for resolving the $A_y(\theta)$ puzzle with 
a 3NF.
All calculations are performed at $E_{Lab}^n$ = 3 MeV, where experimental 
data are available for the differential cross section \cite{Sc83} and 
$A_y(\theta)$ \cite{Mc94}.
The Argonne $V_{18}$ model (AV18) \cite{Wi95} is used as the input NN 
potential throughout this paper.

Our method for calculating the 3N continuum state is based on a natural 
extension of the bound state calculation \cite{Sa86,Is87,Is94}, in which the 
Faddeev equation is expressed as an integral equation in coordinate space.
In the continuum state calculation, there appear additional singularities in 
the Faddeev integral kernel, which are absent in the bound state 
calculation: elastic singularity and three-body breakup singularity.
The latter does not appear at energies below the three-body breakup 
threshold as in the present work.
The former singularity can be easily treated by the usual subtraction 
method \cite{Sa78}.
In the present calculation, 3N partial wave states for which NN and 3N 
forces act, are restricted to those with total two-nucleon 
angular momenta $j \le 2$.
The total 3N angular momenta ($J$) is truncated at $J=19/2$, while 3NF is 
switched off for 3N states with $J\ge9/2$. 
These truncating procedures are known to be valid for the low-energy 
($E_{Lab}^n$ = 3 MeV) n-d scattering.

The two-pion exchange three nucleon potential has the following form in 
momentum space:
\begin{eqnarray}
V({\bf q},{\bf q}')&=& 
  \frac1{(2\pi)^6} \left(\frac{f_\pi}{\mu}\right)^2
  {\frac{F(q^2)}{q^2+\mu^2} \frac{F(q'^2)}{q'^2+\mu^2}}
  (\mbox{\boldmath $\sigma$}_1\cdot{\bf q})
  (\mbox{\boldmath $\sigma$}_2\cdot{\bf q}') 
\nonumber\\
 &\times&
 \Bigl[ ({\vec \tau}_1\cdot{\vec \tau}_2)
     \{a+b({\bf q}\cdot{\bf q}')+c(q^2+q'^2)\}  \Bigr.
\nonumber\\
  &&+ \Bigl. (i{\vec \tau}_3\cdot{\vec \tau}_2\times{\vec \tau}_1)
    (i\mbox{\boldmath $\sigma$}_3\cdot{\bf q}\times{\bf q}') d  \Bigr] ,
\label{eq:2piE3NP}
\end{eqnarray}
where ${\bf q}$ and ${\bf q}'$ are the momenta of the propagating pions, 
$\mu$ is 
the pion mass and  $F(q^2)$ a form factor which is parameterized as 
the dipole form with a cutoff mass $\Lambda$.
The parameters, $a$, $b$, $c$, and $d$, for 
the BR${}^\prime$-3NF \cite{Co83} and
the BR-3NF \cite{Ro86} are shown in Table\ \ref{tab:3NP-param}.
Since the Brazil 3NF model is based on the effective Lagrangian approach, in 
which several diagrams are considered explicitly, we can separate out the 
3NF component which results from $\Delta$-mediated diagram.
The parameters for this 3NF component, which should correspond to the UR-3NF, 
are shown in Table\ \ref{tab:3NP-param} as BR$_{\Delta}$.

%
\begin{table}[t]
\caption{Various parameters for the three-nucleon potentials, 
Eq.\ (\protect\ref{eq:2piE3NP}), used in the present work.}
\label{tab:3NP-param}
\begin{center}
\begin{tabular}{lcccc}
\hline\hline
3NF & $a~(\mu^{-1})$ & $b~(\mu^{-3})$ & $c~(\mu^{-3})$ & 
  $d~(\mu^{-3})$ \\
\hline
BR${}^\prime$ & -1.05 & -2.29 & 0.00 & -0.768 \\
BR & 1.05 & -2.29 & 1.05 & -0.768 \\
BR{$_\Delta$}    &   0.00 & -1.49 &  0.00    & -0.373 \\
\hline\hline
\end{tabular}
\end{center}
\end{table}

The cutoff mass $\Lambda$ is chosen so as to reproduce the triton binding 
energy.
The value of 700 MeV is used for the BR${}^\prime$-3NF and 
the BR-3NF, and 800 MeV for the BR$_\Delta$-3NF.
Hereafter these are designated as 
BR${}^\prime_{700}$, BR$_{700}$, and BR$_{\Delta,800}$, respectively.

\begin{figure}[t]
\begin{center}
\epsfig{file=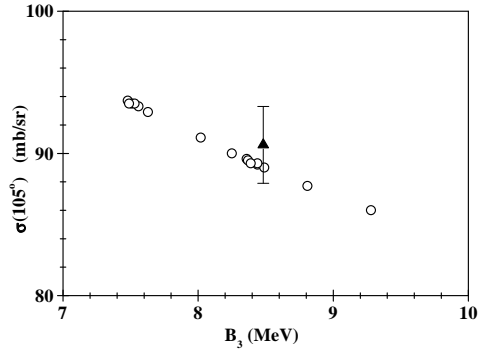, height=5cm}
\end{center}
\caption{Calculated values of the n-d differential cross section at 
$\theta=105^\circ$ for $E_{Lab}^n$ = 3 MeV plotted against the calculated 
triton binding energy $B_3$.
Experimental value is taken from Ref.\ \protect\cite{Sc83}. }
\label{fig:dcs-b3}
\end{figure}

In general, analyzing powers are defined as a difference 
between cross sections with different orientations of incoming particles 
normalized to unpolarized cross sections. 
Therefore, 
before discussing the n-d polarization observables, we make a comment 
on effects of a 3NF on the unpolarized n-d differential cross section (DCS). 
From calculations for various combinations of NN potentials and 3NF models, 
we found that the calculated values of the DCS around the minimum region 
($\theta=105^\circ$) have a correlation with those of the triton binding 
energy, $B_3$.
This is shown in Fig.\ \ref{fig:dcs-b3}, where we plot the calculated values 
of the DCS($105^\circ$) for $E_{Lab}^n$ = 3 MeV against the calculated $B_3$.
The n-d DCS consists of spin-doublet scattering, spin-quartet 
scattering, and their interference terms \cite{Ko86}.
The above correlation can be understood as a result of the well known 
relation 
between the doublet scattering length (${}^2a$) and $B_3$: the Phillips plot.
The $S$-wave DCS at low-energy is proportional to $1/( k^2+1/a^2 )$, 
where $a$ is the scattering length and $k$ is the momentum, 
which means that the DCS decreases if the scattering length $a$ becomes 
smaller.
In fact, the calculated value of the doublet scattering length, 1.35 fm for 
AV18 ($B_3$ = 7.51 MeV), turns out to be 0.68 fm for AV18+BR$_{700}$ 
($B_3$ = 8.36 MeV), while the quartet scattering length is unaffected by a 3NF.
Thus the decrease of the DCS with the introduction of a 3NF should be 
attributed to reproducing the triton binding energy.
In Fig.\ \ref{fig:dcs-b3}, we observe that with reproducing $B_3$, the 
DCS($105^\circ$) gets closer to the central value of the experiment.
However, due to a rather large error bar, it is not conclusive whether the 
decrease is favored.

In Table\ \ref{tab:Obs-3NP}, calculated values of $B_3$; $A_y(105^\circ)$ 
and DCS$(105^\circ)$ at $E_{Lab}^n$ = 3 MeV are shown for AV18, 
AV18+BR${}^\prime_{700}$, AV18+BR$_{700}$, and AV18+BR$_{\Delta,800}$, 
together with the corresponding experimental values, $A_y(\theta)$ at 
$\theta = 104.0^\circ$ \cite{Mc94} and DCS at 
$\theta = 103.9^\circ$ \cite{Sc83}. 
We observe slight decrease (increase) of 
$A_y(105^\circ)$ for AV18+BR$_{700}$ 
(AV18+BR${}^\prime_{700}$ and AV18+BR$_{\Delta,800}$) compared to AV18, 
which is consistent with previous calculations \cite{Wi94,Is94,Ki95,Ki96}.
In Table\ \ref{tab:Obs-3NP}, results for a modified version of AV18 
(Mod-AV18) \cite{To98}, in which factors 0.96, 0.98, and 1.06 multiply 
the ${}^3P_0$, ${}^3P_1$, and ${}^3P_2$ AV18 potentials, respectively, 
are also shown.
The modification causes relatively large effects on NN analyzing power: 
overshooting of peak values of neutron-proton $A_y(\theta)$ by 
about 30 \% and 10 \% for $E_{Lab}^n$ = 3 MeV and 25 MeV, respectively,  
which has been strongly criticized \cite{Hu98}.

\begin{table}[t]
\caption{The results of the triton binding energy; the neutron analyzing 
power and the differential cross section at $\theta=105^\circ$ for the n-d 
scattering at $E_{Lab}^n$ = 3 MeV with AV18 + various 3NF and the modified 
AV18 (Mod-AV18).
Experimental values are $A_y(104.0^\circ)$ \protect\cite{Mc94} and 
$\sigma(103.9^\circ)$ \protect\cite{Sc83}. }
\label{tab:Obs-3NP}
\begin{center}
\begin{tabular}{llll}
\hline\hline
 & $B_3$ (MeV) & $A_y(105^\circ)$ (\%) & $\sigma(105^\circ)$ (mb/sr)\\
\hline
Exp. & 8.48 & 5.96 $\pm$ 0.13  & 90.6 $\pm$ 2.7\\
AV18               & 7.51  & 4.29 & 93.5\\
AV18+BR$^\prime_{700}$ & 8.44  & 4.50 & 89.2\\
AV18+BR$_{700}$        & 8.36  & 3.62 & 89.6\\
AV18+BR$_{\Delta,800}$ & 8.37  & 4.43 & 89.5\\
Mod-AV18           & 7.53  & 5.11 & 93.4\\
\hline\hline
\end{tabular}
\end{center}
\end{table}

The BR-3NF and the BR$_\Delta$-3NF give opposite $A_y(\theta)$ effects. 
From Table\ \ref{tab:3NP-param}, we see that there is no term 
corresponding to the coefficients $a$ and $c$ in the BR$_\Delta$-3NF, 
which comes from the $\pi N$ $S$-wave scattering amplitude. 
Thus it is interesting to see how each term in 2$\pi$E-3NF affects 
$A_y(\theta)$.
To see this, we calculate the n-d scattering at $E_{Lab}^n$ = 3 MeV 
taking into account each term corresponding to the parameter $a$, or $b$, 
or $c$, or $d$ in BR$_{700}$ in addition to AV18.
Each potential is designated as BR$_a$, BR$_b$, BR$_c$, and BR$_d$, 
respectively.
The results are shown in Table\ \ref{tab:BR-abcd}. 
From Table \ref{tab:BR-abcd}, we see that $A_y(105^\circ)$ decreases for 
BR$_a$, BR$_c$, and BR$_d$, but increases for BR$_b$.
Especially BR$_c$ gives a large $A_y(105^\circ)$ effect. 
From this, it is concluded that the contribution from BR$_b$ is larger than 
the one from the other terms in BR${}^\prime$-3NF, 
BR$_{\Delta}$-3NF, and UR-3NF to give a small increase in 
$A_y(\theta)$, while the contribution from BR$_c$ is overwhelming in 
lowering $A_y(\theta)$ in BR-3NF and TM-3NF.
The BR$_c$ includes the so-called contact term, 
which was argued to be excluded to avoid an odd behavior 
of the 3NF at short range \cite{Ro86}, or from a viewpoint of chiral 
constraints \cite{Fr99}. 
It is remarked that BR${}^\prime$-3NF is obtained from BR-3NF 
with a prescription to remove the contact term: 
replacing the coefficient $a$ by $a - 2 \mu^2 c$ and setting $c$ to be 
zero.
Therefore we may express that the different $A_y(\theta)$ effect of 
BR-3NF from that of BR${}^\prime$-3NF arises from the existence of the 
contact term.

\begin{table}[t]
\caption{The results of the triton binding energy; the neutron analyzing 
power and the differential cross section at $\theta=105^\circ$ for the n-d 
scattering at $E_{Lab}^n$ = 3 MeV with AV18 + each term in BR-3NF. }
\label{tab:BR-abcd}
\begin{center}
\begin{tabular}{lccc}
\hline\hline
 & $B_3$ (MeV) & $A_y(105^\circ)$ (\%) & $\sigma(105^\circ)$ (mb/sr)\\
\hline
AV18+BR$_a$ & 7.48 & 4.17 & 93.7\\
AV18+BR$_b$ & 8.25 & 4.55 & 90.0\\
AV18+BR$_c$ & 7.56 & 3.81 & 93.3\\
AV18+BR$_d$ & 7.63 & 4.14 & 92.9\\
\hline\hline
\end{tabular}
\end{center}
\end{table}

In the AV18+BR${}^\prime_{700}$ (AV18+BR$_{\Delta,800}$) calculation, the 
DCS$(105^\circ)$ decreases by 5 \%  (4 \%) compared to the AV18 calculation, 
while $A_y(105^\circ)$ increases by 5 \% (3 \%).
On the other hand, in the Mod-AV18 calculation, $A_y(105^\circ)$ increases 
with little change in the DCS$(105^\circ)$.
Thus there is an essential difference between effects on $A_y(\theta)$ from 
the 2$\pi$E-3NF and from the modification of the ${}^3P_J$ NN force.
The former is an increase in $A_y(\theta)$ simply due to the decrease of the 
DCS due to the effect of reproducing the triton binding energy.

\begin{figure}[t]
\begin{center}
\epsfig{file=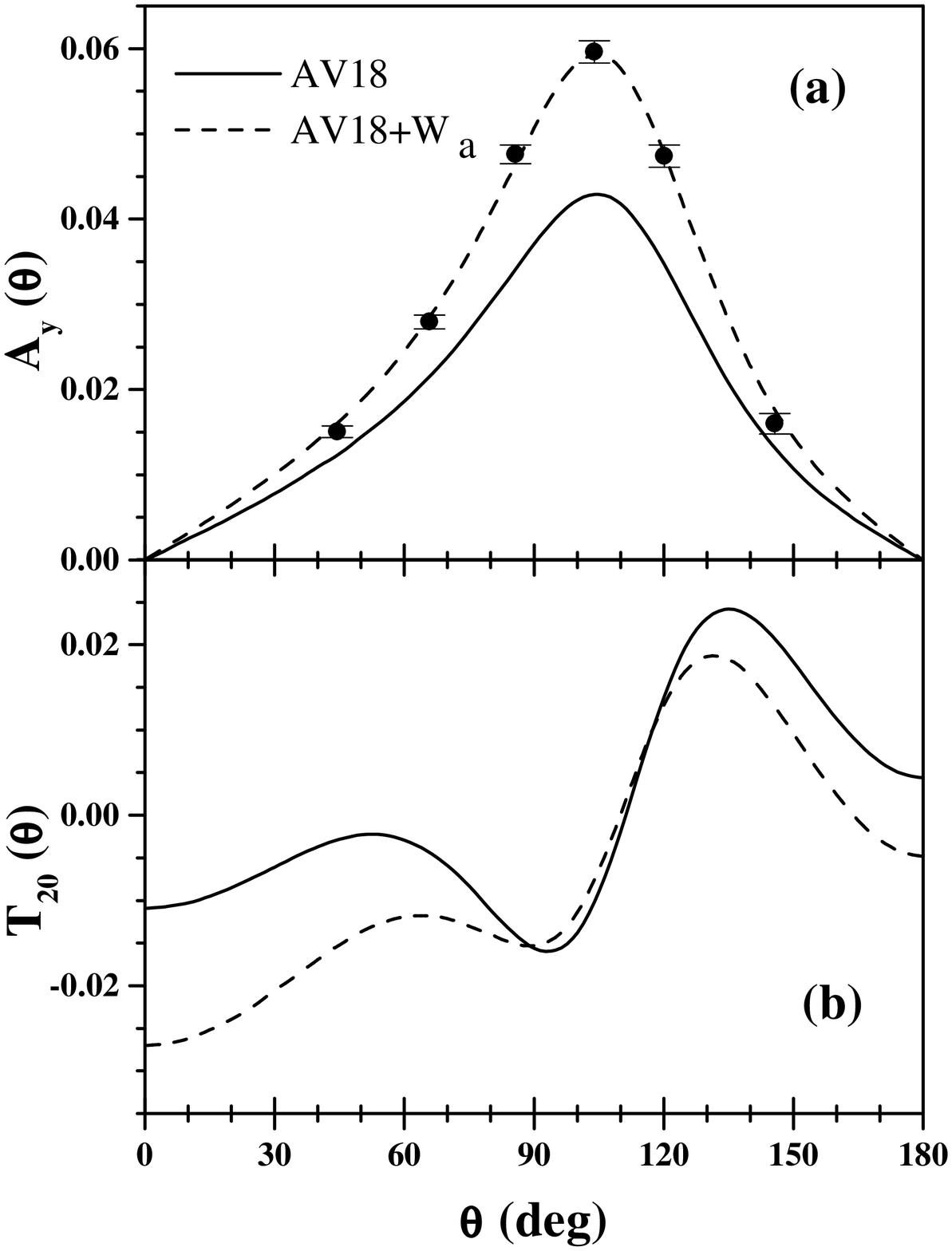, height=8cm}
\end{center}
\caption{$A_y(\theta)$ at $E_{Lab}^n$ = 3 MeV (a) and $T_{20}(\theta)$ at 
$E_{Lab}^d$ = 6 MeV (b) calculated with AV18 (solid lines) and AV18+$W_a$ 
(dashed lines).
Experimental data of $A_y(\theta)$ are taken from Ref.\ \protect\cite{Mc94}.}
\label{fig:Wa-nd}
\end{figure}

In Table\ \ref{tab:BR-abcd}, we observe that each term in BR-3NF gives 
quite different effects in $A_y(\theta)$.
Next, we investigate each effect of the four terms in the 2$\pi$E-3NF.
To do so, we introduced only the $a$ ($b$, $c$, $d$)-term as a 3NF by 
varying the coefficient of $a$ ($b$, $c$, $d$) to reproduce the triton 
binding energy.
These 3NF models are designated as $W_a$, $W_b$, $W_c$, and $W_d$.
The results are shown in Table\ \ref{tab:abcd} together with the values 
of the coefficients.
Although the change of sign in $a$ and $c$, and of the magnitude in $a$ 
compared to the original values in Table\ \ref{tab:3NP-param} may be 
unnatural, these 3NF models might be useful as phenomenological ones
which reproduce the triton binding energy within a restricted functional 
form.
A variety of $A_y(\theta)$ effects are observed from these 3NF models: 
a large increase due to $W_a$; a small increase due to $W_b$ and $W_c$; 
a relatively large decrease due to $W_d$, besides the decrease of in the 
DCS$(105^\circ)$ due to the binding energy effect.
It is remarkable that $W_a$ seems to reproduce the experimental value of 
$A_y(105^\circ)$ quite well.
However, it turns out that the deuteron tensor analyzing powers are modified 
improperly by $W_a$ at the same time.
In Fig.\ \ref{fig:Wa-nd}, we plot $A_y(\theta)$ at $E_{Lab}^n$ = 3 MeV (a) 
and $T_{20}(\theta)$ at $E_{Lab}^d$ = 6 MeV (b) calculated with AV18 
(solid lines) and AV18+$W_a$ (dashed lines).
We see that the experimental data for $A_y(\theta)$ are well reproduced with 
the introduction of $W_a$, and $T_{20}(\theta)$ is significantly modified.
Although there is no $T_{20}(\theta)$ data for n-d scattering, 
recent precise measurements of tensor analyzing powers for 
proton-deuteron scattering are reported to be well reproduced by 
calculations without a 3NF \cite{Sh95}.
Thus such distortion of $T_{20}(\theta)$ for n-d scattering may produce 
another "puzzle".

\begin{table}[t]
\caption{The results of the triton binding energy; the neutron analyzing 
power and the differential cross section at $\theta=105^\circ$ for the n-d 
scattering at $E_{Lab}^n$ = 3 MeV with AV18 + $W_a$, $W_b$, $W_c$ and $W_d$.}
\label{tab:abcd}
\begin{center}
\begin{tabular}{lccc}
\hline\hline
 & $B_3$ (MeV) & $A_y(105^\circ)$ (\%) & $\sigma(105^\circ)$ (mb/sr)\\
\hline
AV18+$W_a$ & 8.49 & 5.93 & 89.0\\
 $(a=-14.4 \mu^{-1})$ \\
AV18+$W_b$ & 8.50 & 4.64 & 89.0\\
$(b=-2.90 \mu^{-3})$ \\
AV18+$W_c$ & 8.49 & 5.13 & 88.8\\
$(c=-1.25 \mu^{-3})$ \\
AV18+$W_d$ & 8.50 & 3.65 & 88.9\\
$(d=-3.10 \mu^{-3})$ \\
\hline\hline
\end{tabular}
\end{center}
\end{table}

It is found that the $W_a$-3NF gives a different effect on n-d polarization 
observables than other 3NF models.
We remark that this difference can be seen in the n-d quartet 
$P$-wave phase-shifts: $\delta_{{}^4P_{1/2}}$, $\delta_{{}^4P_{3/2}}$, and 
$\delta_{{}^4P_{5/2}}$, to which $A_y(\theta)$ is known to be sensitive.
The relation between the n-d phase shifts and $A_y(\theta)$ is quite 
complicated, but as derived in the Appendix, a combination 
\begin{equation}
 - 4 M_{{}^4P_{1/2}} - 5 M_{{}^4P_{3/2}} + 9M_{{}^4P_{5/2}}
\label{eq:ay-amp}
\end{equation}
appears in an expression for $A_y(\theta)$,
where $M_{{}^4P_J} = 
  \exp\left(i\delta_{{}^4P_J}\right) \sin\left( \delta_{{}^4P_J} \right)$.
For small phase shift differences, Eq.\ (\ref{eq:ay-amp}) is proportional to
\begin{equation}
4 \Delta_{3/2-1/2} + 9 \Delta_{5/2-3/2} ,
\label{eq:ay-Delta}
\end{equation}
where
$\Delta_{J-J^\prime} = \delta_{{}^4P_{J}}-\delta_{{}^4P_{J^\prime}}$.
Eq.\ (\ref{eq:ay-Delta}) is a convenient expression for $A_y(\theta)$ 
being consistent with 
results of three-nucleon phase shift analysis \cite{To98}.
In Table\ \ref{tab:phase}, we list the calculated values of 
$\delta_{{}^4P_{1/2}}$, $\Delta_{3/2-1/2}$, and $\Delta_{5/2-3/2}$
for some  models presented in this work.
From Table\ \ref{tab:phase} we see that 
$\Delta_{5/2-3/2} \sim 0$
for most cases except for $W_a$, for which 
$\Delta_{3/2-1/2} \sim 0$.
Therefore, 
$A_y(\theta)$ is proportional to $9 \Delta_{5/2-3/2}$ 
($4 \Delta_{3/2-1/2}$) for $W_a$ (the other 3NF models).
The difference of the factors, 9 and 4, explains the reason why $W_a$ gives 
a large increase in $A_y(\theta)$ in spite of the same order of the phase 
shift differences, $\Delta_{5/2-3/2}$ and $\Delta_{3/2-1/2}$.
However, the difference seems to affect incorrectly the deuteron tensor 
analyzing powers.

\begin{table}[t]
\caption{Phase shift for the n-d $^4P_{1/2}$ state and the differences 
$\Delta_{3/2-1/2}$ and $\Delta_{5/2-3/2}$,  which are defined in 
the text, at $E_{Lab}^n$ = 3 MeV. }
\label{tab:phase}
\begin{center}
\begin{tabular}{lccc}
\hline\hline
 & $\delta_{{}^4P_{1/2}}$ & $\Delta_{3/2-1/2}$ & $\Delta_{5/2-3/2}$ \\
\hline
AV18                    & 24.2 & 1.9 & 0.1 \\
AV18+BR$_{700}$         & 24.5 & 1.9 & -0.2\\
AV18+BR$^\prime_{700}$  & 24.6 & 1.4 & 0.4 \\
AV18+BR$_{\Delta,800}$  & 24.5 & 1.6 & 0.3 \\
Mod-AV18                & 24.0 & 2.2 & 0.2 \\
AV18+$W_a$              & 24.9 & 0.2 & 1.4 \\
\hline\hline
\end{tabular}
\end{center}
\end{table}

In summary, we have studied the effects of the 2$\pi$E-3NF, and its 
variations, on some observables for n-d elastic scattering at low energy.
We found that a contact term included in the 2$\pi$E-3NF gives a rather large 
$A_y(\theta)$ effect.
This is the reason why effects on $A_y(\theta)$ by BR/TM-3NF are different 
from those by BR${}^\prime$/UR-3NF, which does not include the contact term.
$A_y(\theta)$ increases by about 5\% with a 2$\pi$E-3NF model in which the 
contact term is eliminated.
However, this increase is essentially the result of a decrease in the 
differential cross section caused by reproducing the triton binding energy.
This contrasts with the increase of $A_y(\theta)$ by the modification of 
the ${}^3P_J$ NN force, which is caused by a variation in spin-dependent 
cross sections. 
We found a phenomenological 3NF model which reproduces both of the triton 
binding energy and $A_y(\theta)$. 
This 3NF originates from $\pi N$ $S$-wave scattering in the intermediate 
state with the strength parameter adjusted to reproduce the triton binding 
energy.
But this 3NF destroys the good fit of the tensor analyzing power at the 
same time.
Since forces arising from the exchange of pions should have a tensor 
character, it seems natural that such forces affect not only the 
spin vector observables but also the spin tensor observables.
A 3NF involving any mechanism other than 2$\pi$E, which might have a 
character of a spin-orbit forces 
as suggested from the modification of the ${}^3P_J$ NN force, 
should be examined to resolve the $A_y(\theta)$ puzzle.

\appendix
\section*{Appendix}


In the spherical base, $A_y(\theta)$ is given by 
\begin{equation}
I(\theta) A_y(\theta) = 
  i I(\theta) \left( T_{+1}(\theta) + T_{-1}(\theta) \right) / \sqrt{2} 
\label{eq:Ay-T}
\end{equation}
with
$I(\theta) = Tr \left( M  M^\dagger \right)$,
and $I(\theta)T_{\kappa}(\theta) = Tr \left( M \tau^1_{\kappa} M^\dagger \right)$,
where $M$ is a transition matrix and $\tau^1_{\kappa}$ is a nucleon rank-1 
spin operator, whose matrix elements 
in the channel-spin representation are  
\begin{eqnarray}
 < s \nu | \tau^1_\kappa | s^\prime \nu^\prime >
&&= (-1)^{2s+\frac12-\nu^\prime}  \sqrt{2} \hat{s}\hat{s}^\prime 
\nonumber\\
\times &&
( s  s^\prime -\nu \nu^\prime | 1 -\kappa )
\left\{\begin{array}{ccc}
 s & s^\prime & 1\\
1/2 & 1/2 & 1
\end{array} \right\} .
\end{eqnarray}
Here, $\hat{n} = \sqrt{2n+1}$, and $s$ is the channel-spin.

With partial-wave amplitudes, $M^{J}_{s \ell s^\prime \ell^\prime}$, 
the transition matrix elements, $M_{s \nu s^\prime \nu^\prime}$, 
are given as \cite{Se69}
\begin{eqnarray}
M_{s \nu s^\prime \nu^\prime}(\theta) &&=
  \sum_{J,\ell,\ell^\prime, m_\ell} \hat{\ell^\prime}
  ( s \ell \nu m_\ell | J \nu^\prime )
\nonumber\\
\times &&  ( s^\prime \ell^\prime  \nu^\prime 0 | J \nu^\prime ) 
  M^{J}_{s \ell s^\prime \ell^\prime}
  Y_\ell^{m_\ell}(\theta,0) .
\end{eqnarray}

Here, we apply some assumptions.

{\em A.} 
We assume off-diagonal matrix elements of the partial wave amplitude to 
vanish:
\begin{equation}
  M^{J}_{s \ell s^\prime \ell^\prime}
= \delta_{s,s^\prime} \delta_{\ell, \ell^\prime}
  M_{{}^{2s+1}\ell_{J}}
\end{equation}

{\em B.} 
Since we are interested in ${}^4P_J$ waves, we consider contribution 
only from $s=3/2$.
Then we have:
\begin{eqnarray}
I(\theta) T_{\kappa}(\theta) 
&&= 4 \sum_{\nu,\nu^\prime,\nu^{\prime\prime}}  (-1)^{-1/2-\nu^\prime} 
  M_{\nu \nu^\prime} M^*_{\nu \nu^{\prime\prime}}
\nonumber\\
&& \times  
  (\frac32 \frac32 -\nu^\prime \nu^{\prime\prime} | 1 -\kappa )
\left\{\begin{array}{ccc}
3/2 & 3/2 & 1\\
1/2 & 1/2 & 1
\end{array} \right\}
\label{eq:IT32}
\end{eqnarray}
with
\begin{eqnarray}
M_{\nu \nu^\prime}(\theta) 
&=& 
  \sum_{J, \ell, m_\ell} \hat{\ell}
  ( \frac32 \ell \nu m_\ell | J \nu^\prime )
 ( \frac32 \ell \nu^\prime 0 | J \nu^\prime ) 
\nonumber\\
&& \times  M_{{}^4\ell_J} Y_\ell^{m_\ell}(\theta,0) 
\label{eq:Mmumus}
\end{eqnarray}
\begin{eqnarray}
M^*_{\nu \nu^{\prime\prime}}(\theta) 
&=& 
  \sum_{J^\prime, \ell^\prime, m_\ell^\prime} \hat{\ell^\prime}
  ( \frac32 \ell^\prime \nu m_\ell^\prime | J^\prime \nu^{\prime\prime} )
  ( \frac32 \ell^\prime \nu^{\prime\prime} 0 | J \nu^{\prime\prime} ) 
\nonumber\\
&& \times
  M^*_{{}^4{\ell^\prime}_J} Y_{\ell^\prime}^{m_\ell^\prime *}(\theta,0) 
\label{eq:Mmumuss}
\end{eqnarray}
{\em C.} 
It is remarked that the shape of $I(\theta) A_y(\theta)$ for the n-d 
scattering is roughly given by $\sin\theta$. 
This $\theta$-dependence arises 
when $(\ell, m_\ell, \ell^\prime, m_\ell^\prime) = (1, \pm 1, 0, 0)$, 
or $(0, 0, 1, \pm 1)$.
For these cases, after evaluating the summation over $\nu$, $\nu^\prime$ 
and $m_\ell$ in Eq.\ (\ref{eq:IT32}), we obtain 
\begin{eqnarray}
I(\theta) T_\kappa(\theta) &\propto&  \kappa Y_1^{\kappa}(\theta,0) 
   M^*_{{}^4S_{1/2}} 
   \sum_{J} M _{{}^4P_J} 
\nonumber\\
& & \times (-1)^{J-1/2} \hat{J}^2
\left\{\begin{array}{ccc}
 1 & 1 & 1\\
3/2 & 3/2 & J
\end{array} \right\}
\label{eq:ITk}
\end{eqnarray}
The summation in Eq.\ (\ref{eq:ITk}) 
is proportional to Eq.\ (\ref{eq:ay-amp})



\begin{thebibliography}{99}
\bibitem{Gl96} 
W. Gl\"ockle, H. Wita\l a, D. H\"uber, H. Kamada, and J. Golak, 
Phys. Rep. {\bf 274}, 107 (1996).

\bibitem{Ko87}
Y. Koike and J. Haidenbauer, Nucl. Phys. {\bf A463}, 365c (1987).

\bibitem{Wi88}
H. Wita\l a, W. Gl\"ockle, and T. Cornelius, 
Nucl. Phys. {\bf A491}, 157 (1988).

\bibitem{Mc94}
J. E. McAninch, L. O. Lamm, and W. Haeberli, 
Phys. Rev. C {\bf 50}, 589 (1994).

\bibitem{St94}
V. G. J. Stokes,  R. A. M. Klomp, C. P. F. Terheggen, and 
J. J. de Swart, Phys. Rev. C {\bf 49}, 2950 (1994).

\bibitem{Wi95} 
R. B. Wiringa, V. G. J. Stokes, and R. Schiavilla, 
Phys. Rev. C {\bf 51}, 38 (1995).

\bibitem{Ma96}
R. Machleidt, F. Sammarruca, and Y. Song,
Phys. Rev. C {\bf 53}, R1483 (1996).

\bibitem{Wi91} 
H. Wita\l a and W. Gl\"ockle, Nucl. Phys. {\bf A528}, 48 (1991).

\bibitem{Ta91} 
T. Takemiya, Prog. Theor. Phys. {\bf 86}, 975 (1991).

\bibitem{To98}
W. Tornow, H. Wita\l a, and A. Kievsky, Phys. Rev. C {\bf 57}, 555 (1998).

\bibitem{Do98}
P. Doleschall, Few-Body Syst. {\bf 23}, 149 (1998).

\bibitem{Wi92} 
H. Wita\l a, W. Gl\"ockle and T. Takemiya, 
Prog. Theor. Phys. {\bf 88}, 1015 (1992).

\bibitem{Hu98}
D. H\"uber and J. L. Friar, Phys. Rev. C {\bf 58}, 674 (1998).

\bibitem{To98b}
W. Tornow and H. Wita\l a, Nucl. Phys. {\bf A637}, 280 (1998).

\bibitem{Co79}
S. A. Coon, M. D. Scadron, P. C. McNamee, B. R. Barrett, D. W. E. Blatt, 
and B. H. J. McKellar, Nucl. Phys. {\bf A317}, 242 (1979);
S. A. Coon and W. Gl\"ockle, Phys. Rev. C {\bf 23}, 1790 (1981).

\bibitem{Co83} 
H. T. Coelho, T. K. Das, and M. R. Robilotta, 
Phys. Rev. C {\bf 28}, 1812 (1983).

\bibitem{Ro86}
M. R. Robilotta and H. T. Coelho, Nucl. Phys. {\bf A460}, 645 (1986).

\bibitem{Wi94} 
H. Wita\l a, D. H\"uber, and W. Gl\"ockle, Phys. Rev. C {\bf 49}, R14 (1994). 

\bibitem{Is94}
S. Ishikawa, Y. Wu, and T. Sasakawa, AIP Conf. Proc. {\bf 334}, 840 (1994).

\bibitem{Ki95}
A. Kievsky, M. Viviani, and S. Rosati, Phys. Rev. C {\bf 52}, R15 (1995). 

\bibitem{Ki96} 
A. Kievsky, S. Rosati, W. Tornow, M. Viviani, 
Nucl Phys. {\bf A607}, 402 (1996).

\bibitem{Fu57}
J. Fujita and H. Miyazawa, Prog. Theor. Phys. {\bf 17}, 360 (1957).

\bibitem{Sc83} 
P. Schwarz, H. O. Klages, P. Doll, B. Haesner, J. Wilczynski, B. Zeitnitz, 
and J. Kecskemeti, Nucl. Phys. {\bf A398}, 1 (1983).

\bibitem{Sa86}
T. Sasakawa and S. Ishikawa, Few-Body Syst. {\bf 1}, 3 (1986).

\bibitem{Is87}
S. Ishikawa, Nucl. Phys. {\bf A463}, 145c (1987).

\bibitem{Sa78}
T. Sasakawa, Phys. Rev. C {\bf17}, 2015 (1978).

\bibitem{Ko86}
Y. Koike and Y. Taniguchi, Few-Body Syst. {\bf 1}, 13 (1986).

\bibitem{Fr99}
J. L. Friar, D. H\"uber, and U. van Kolck, Phys. Rev. C {\bf 59}, 53 (1999).

\bibitem{Sh95}
S. Shimizu, K. Sagara, H. Nakamura, K. Maeda, T. Miwa, N. Nishimori, 
S. Ueno, T. Nakashima, and S. Morinobu,
Phys. Rev. C {\bf 52}, 1193 (1995).

\bibitem{Se69} 
R. G. Seyler, Nucl. Phys. {\bf A124}, 253 (1969).

\end{thebibliography}
\end{document}